\documentstyle[12pt]{article}

\parskip=\medskipamount 
\def\be{\begin{equation}}
\def\bea{\begin{eqnarray}}
\def\nn{\nonumber}
\def\ee{\end{equation}}
\def\eea{\end{eqnarray}}

\def\CR{\hbox{{$\cal R$}}} 
 
\def\CM{\hbox{{$\cal M$}}} \def\CP{\hbox{{$\cal P$}}}

\def\sgn{{\rm sgn}}

\begin{document}
\begin{titlepage}

\begin{center}
{\Large \bf Example of q-deformed Field Theory}\\
\vspace{2.cm}
{\normalsize \bf V.Bardek, M.Dore\v si\' c and S.Meljanac\footnote
{e-mail address: bardek@thphys.irb.hr\\
 e-mail address: doresic@thphys.irb.hr\\
 e-mail address: meljanac@thphys.irb.hr}}\\
\vspace{0.5cm}
Department of Theoretical Physics \\
Rudjer Bo\v skovi\' c Institute, P.O.B. 1016,\\
41001 Zagreb,CROATIA \\
\vspace{2cm}
{\large \bf Abstract}
\end{center}
\vspace{0.5cm}
\baselineskip=24pt
The non-relativistic Chern-Simons theory with the single-valued
anyonic field is proposed as an example of q-deformed field theory.
The corresponding q-deformed algebra
interpolating between bosons and fermions,both in position and
momentum spaces, is analyzed.A possible generalization to a space
with an arbitrary dimension is suggested.

\vspace{30mm}

\begin{center}
\end{center}
\end{titlepage}

\setcounter{page}{1}
\newpage
\baselineskip=24pt
\vspace{1.5cm}     
In the last few years considerable interest has been devoted
to non-commutative geometry~ \cite{con} and quantum groups
{}~ \cite{fad} ,both in mathematics and physics.The mathematical
knowledge is rapidly growing up.On the other hand,the role
of these remarkable mathematical structures in physics is still
not clear.Moreover,their connection with fundamental physical
concepts and laws is missing.

Anyonic physics ~ \cite{sem} in the Chern-Simons description can
be viewed as a particular realization of these ideas,although
in ~2+1~ dimensions only.As was shown in ~ \cite{dor}, the anyonic
graded-commutation algebra represents a typical q-deformed
structure ~ \cite{gre} ,with ~q~ being a unimodular complex
parameter,determining the type of intermediate anyonic statistics.
For consistency of this approach,it is essential that anyons
are multi-valued objects ~ \cite{sem}.

In this paper we present another type of q-deformed field
theory that can be generalized in different directions.We
start with the non-relativistic Chern-Simons theory.Using
the single-valued transformation we obtain a different type of
anyonic field.The corresponding graded-commutation relations for
this anyonic field are a generalization of those for quantum-group
particles ~ \cite{bor}.We point out that our anyonic field is
a single-valued object defined on the ordinary ~2+1~ space-time.
This field is not free and the corresponding commutation relations
in momentum space are non-local.Furthemore,we construct
{}~$\cal R$~-matrices for anyon operators both in position and
momentum space and show that they satisfy the Yang-Baxter equation
and the Hecke condition.
Finally,we investigate the origin
of the transmutation from bosons to anyons ( or fermions to
anyons) and suggest a generalization to a space with an arbitrary
number of dimensions.

The quantum-mechanical problem of non-relativistic bosons interacting
with the U(1) gauge field described by the Chern-Simons Lagrangian has
been formulated and solved by Jackiw and Pi ~ \cite{jack}.They have
shown that the corresponding Hamiltonian can be reduced to the
following form:

\begin{eqnarray}
 H & = & \frac{1}{2} \int d^2 r \; [ ( \vec \nabla - i \:
 \vec A( \vec r \, )) \: \Psi( \vec r \, ) ]^{\dagger} \:
 [ ( \vec \nabla - i \:
 \vec A( \vec r \, )) \: \Psi( \vec r \, ) ] \; .           \label{1}
\end{eqnarray}

The quantum field operator  $\Psi( \vec r \, )$ and its hermitian
conjugate $\Psi^{\dagger}( \vec r \, )$ obey the bosonic commutation
relations at equal times:
\begin{eqnarray}
[ \; \Psi ( \vec r \, ) \; , \; \Psi ( \vec r \, ' \, ) \; ] &=& 0
\; ,
 \nonumber\\
{[ \; \Psi( \vec r \, ) \; , \; {\Psi}^{\dagger} ( \vec r \, ' \, )
 \; ]}
 &=& { \delta \, ( \, \vec r - \vec r \, ' \, ) \; .}       \label{2}
\end{eqnarray}

The vector potential ~$\vec A$~ can be completely described by the
bosonic-number density operator ~$\rho$ ~ :

\begin{eqnarray}
\vec A \: ( \vec r \, ) \; &=& \; - \lambda \: \hat{n}  \: {\times} \:
 {\int} \frac{ \vec r - \vec r \, '}{{\mid \vec r - \vec r \, '
 \mid}^2} \; \rho( \vec r \, ' \, ) \; d^2 r' \; , \\       \label{3}
\rho( \vec r \,) \; &=& \; {\Psi}^{\dagger}({\vec r} \, ) \:
\Psi ({\vec r} \, ) \; ,                                    \label{4}
\end{eqnarray}

where ~ $\lambda$ ~ is the so-called statistical parameter,and
{}~$\hat{n}$~ is the unit vector orthogonal to the plane.
In other words,the gauge field has no independent dynamics.

Now we introduce a new field ~$\tilde{\Psi} ( \vec r \, )$~
by performing unitary transformation,i.e. by the phase
redefinition of the bosonic field  ~$\Psi ( \vec r \, )$~:

\begin{equation}
\tilde{\Psi} ( \vec r \, ) \; = \; e^{-i \, \omega ( \vec r \, )} \:
\Psi ( \vec r \, ) \; ,                                     \label{5}
\end{equation}

where

\begin{equation}
\omega ( \vec r \, ) \; = \; - \lambda \int d^2 r' \; \theta
( \, \vec r - \vec r \, ' \, ) \rho ( \vec r \, ' \, ) \; . \label{6}
\end{equation}
Here ~$\theta( \vec r )$~ is the polar angle of the radius vector
{}~$\vec r$~ in respect to the positive x-axis; it is defined
as a single-valued function and with
{}~$0 \leq \theta ( \vec r ) < 2 \pi $~.In this way we
have introduced
a cut along the positive x-axis,which induces additional,residual
interactions.Then we find that

\be
\vec \nabla \omega ( \vec r \,) \; = \vec A( \vec r \,) -
\vec A_{res}( \vec r \, ) \, ,                              \label{7}
\ee
where  the residual vector potential is given by
\be
\vec A_{res}( \vec r \, ) = 2 \pi \lambda \: \vec j
\int_{x}^{\infty}dx' \rho (x',y) \; .                       \label{8}
\ee
Here ~$\vec i , \vec j $~ are unit vectors along the positive
x- and y-axis,respectively.In this way we obtain the following
Hamiltonian for the new field ~$\tilde{\Psi} ( \vec r \, )$~:

\bea
 H & = & \frac{1}{2} \int d^2 r \; [ ( \vec \nabla - i \:
\vec A_{res}( \vec r \, )) \: \tilde \Psi( \vec r \, ) ]^{\dagger} \:
 [ ( \vec \nabla - i \:
 \vec A_{res}( \vec r \, )) \: \tilde \Psi( \vec r \, ) ] \; .
                                                            \label{9}
\eea
Note that
{}~$\vec B = \vec \nabla \times \vec A_{res} = 2 \pi \, \lambda \,
\hat n \, \rho ( \vec r \, )$~, i.e. that the magnetic field
{}~$\vec B$~does not change under the transformation~(\ref{5})~.

The commutation relations~(\ref{2})~written in terms of the new
field ~$\tilde \Psi ( \vec r \, )$~ are now modified as

\bea
\tilde{\Psi}(\vec r \, ) \: \tilde{\Psi}(\vec r \, ' \, )
\; - \; e^{-i \, \lambda \, \Delta( \vec r - \vec r \, ' \, )}
\tilde{\Psi}(\vec r \, ' \, ) \: \tilde{\Psi}(\vec r \, )
&=& 0 \; ,
\nn\\
\tilde{\Psi}(\vec r \, ) \: \tilde{\Psi}^{\dagger}(\vec r \, ' \, )
\; - \; e^{i \, \lambda \, \Delta( \vec r - \vec r \, ' \, )}
\tilde{\Psi}^{\dagger}(\vec r \, ' \, ) \: \tilde{\Psi}(\vec r \, )
 & = & \delta \,( \, \vec r - \vec r \, ' \, ) \; ,         \label{10}
\eea

where ~$\Delta$~ denotes the difference

\bea
\Delta \,( \, \vec r - \vec r \, ' \, ) \; &=& \;
\theta \, ( \, \vec r - \vec r \,' \, ) \; - \;
\theta \, ( \, \vec r \, ' - \vec r \, ) \nn \\
&=& \left \{ \begin{array}{ll}
             - \pi \sgn (y-y')    & {\rm if} \; \; y-y' \neq 0  \\
             - \pi \sgn (x-x')    & {\rm if} \; \; y-y' = 0 \; .
                                                            \label{11}
             \end{array}
    \right.
\eea

We point out that the antisymmetry of the ~$\Delta ( \vec r \, )$~
function is a necessary and sufficient condition for the consistency
of the above graded-commutation relations.In a multi-valued
picture , ~$\Delta$~ is a multi-valued constant,i.e. \\
{}~$\Delta = \{ \pi ( 1+2z ), z \in Z \}$~,satisfying the antisymmetry
condition in the sense of the set equality ~$\Delta = - \Delta$~.

In the above approach, the grading factor is not a multi-valued
constant but a single-valued function of difference of the
appropriate coordinates.

We interpret the ~$\tilde \Psi ( \vec r \, )$~ field as a new
type of anyonic field,or the q-deformed field ( quantum-group field).
This new field is a generalization of the finite-dimensional
quantum spaces with coordinates
{}~$\psi_{1}, \, \bar \psi_{1}, \dots \psi_{n}, \, \bar \psi_{n}$~
satisfying the relations \cite{bor}

\bea
\begin{array}{ccc}
 \psi_{i} \bar \psi_{i} = \bar \psi_{i} \psi_{i}, &
 1 \leq i \leq n, & \\
 \psi_{i} \psi_{j} = q \psi_{j} \psi_{i} \, , &
 \bar \psi_{i} \bar \psi_{j} = q \bar \psi_{j} \bar \psi_{i} & \\
 \psi_{i} \bar \psi_{j} = q^{-1} \bar \psi_{j} \psi_{i} \, , &
 \bar \psi_{i} \psi_{j} = q \psi_{j} \bar \psi_{i} \, , &
 1 \leq i < j \leq n , \\
 \mid q \mid =1 , q^{m} = 1 .                               \label{12}
\end{array}
\eea
In this way, finite-dimensional quantum spaces are generalized
to infinite-dimensional quantum spaces by the graded-
commutation relations~(\ref{10})~ . Hence, formally, we have
\be
\psi_{i} \rightarrow \tilde \Psi ( \vec r \, ) , \hspace{3.5cm}
\psi_{j} \rightarrow \tilde \Psi ( \vec r \, ' \, )         \label{13}
\ee
and
\be
q \rightarrow e^{-i \lambda \pi} , \hspace{2cm}
- \pi \sgn (i-j) \rightarrow \Delta ( \vec r - \vec r \, ' \, ) \: .
                                                            \label{14}
\ee
The structure of the ~2+1~ space-time is not changed.If
{}~$\vec r = \vec r \, ' \,$~,the field ~$\tilde \Psi ( \vec r \, )$~
behaves as a bosonic field,whereas for
{}~$\vec r \neq \vec r \, ' \,$~,it behaves as a q-deformed field
obeying fractional statistics for ~$q \neq \pm 1$~.The graded
algebra~(\ref{10})~is different from the quon algebra
 ~ \cite{dor,gre}.

We note that if we had chosen the fermionic field instead of the
bosonic field ~$\Psi$~ , as a generic field for anyons in
Eq.~(\ref{5})~,
our q-commutator algebra~(\ref{10})~ would have changed the relative
sign
between bilinears,transforming q-commutators into q-anticommutators.
An immediate consequence is the appearance of the hard-
core condition for any value of ~$\lambda$~,including even the
bosonic case ~ \cite{sci}.This can be understood as a counterpart
to the absence of any hard-core condition (even for fermions,
{}~$\lambda = 1$~ ) in the bosonic-based approach.Hence,in both
approaches,the continuous interpolation between ordinary bosons
and fermions is not fully achieved.In fact, the anyonic field,
Eqs.~(\ref{5}) and (\ref{6})~, with ~$\lambda = 1$~,is the same as the
continuum limit of the set of oscillators in Green's ansatz for
para-Bose statistics ~\cite{ohn}.

It is interesting to note that both approaches can be unified
into one for\\
{}~$ \vec r \neq \vec r \, '$~.Then
\be
e^{i \lambda \Delta} =  \cos \lambda \pi \:
+ \: i \, \frac{\Delta}{\pi} \, \sin \lambda \pi \; .           \label{15}
\ee
The substitution
{}~$\lambda = \lambda ' \, + \, 1 \; mod \, 2$~ leads to the
q-fermionic algebra.This can be viewed as a complexification
of the quon algebra of Greenberg~ \cite{gre}.

Operating on the vacuum,the anyonic operators
{}~$\tilde{\Psi}^{\dagger}(\vec r)$~ create a single-valued
quantum state.For example, a two-particle state is given by
\bea
\Psi(\vec r_{1},\vec r_{2}) &=&
\tilde{\Psi}^{\dagger}(\vec r_{1}) \:
\tilde{\Psi}^{\dagger}(\vec r_{2}) \; |0>
\nn\\
&=& e^{-i \, \lambda \, \theta( \vec r_{1} - \vec r_{2} )} \:
{\Psi}^{\dagger} (\vec r_{1} ) \: {\Psi}^{\dagger}(\vec r_{2})
 \; |0>.                                                    \label{37}
\eea
We note that this state changes phase by ~$\pm \lambda \pi$~
when two particles are interchanged,depending upon the rotation
being clockwise or anticlockwise.The N-particle state can be
constructed analogously:
\be
\Psi(\vec r_{1},\vec r_{2}, \dots , \vec r_{N}) =
 e^{-i \, \lambda \, {\Sigma}_{i<j} \theta( \vec r_{i} - \vec r_{j} )}
\: {\Psi}^{\dagger} (\vec r_{1} ) \: {\Psi}^{\dagger}(\vec r_{2})
\dots {\Psi}^{\dagger} (\vec r_{N} )
 \; |0>.                                                    \label{38}
\ee
The single-valued phase factors in ~(\ref{37})~or~(\ref{38})
are a one-dimensional representation of the braid group ~\cite{wu}.

Using the q-commutation algebra ~(\ref{10})~
it is easy to verify that all anyonic states in position
space have a non-negative squared norm. \\

For completeness , we want to formulate and analyze the q-deformed
algebra~(\ref{10})~ in momentum space.Hence,we should Fourier transform
the non-relativistic fields ~$\Psi ( \vec r \, )$~ and
{}~$\tilde \Psi ( \vec r \, )$~.The corresponding annihilation
operators are
\bea
a( \vec k \,) &=& \frac{1}{2 \pi} \int d^2 r \: e^{-i \vec k \vec r}
\Psi( \vec r \, ) \; ,
\nn\\
\tilde a ( \vec k \,) &=& \frac{1}{2 \pi} \int d^2 r \:
e^{-i \vec k \vec r} \tilde \Psi ( \vec r \, ) \; .         \label{16}
\eea

In terms of these operators, the q-deformed algebra ~(\ref{10})~
translates into

\bea
\tilde a( \vec p \, ) \tilde a( \vec q \, ) \; - \;
 \frac{1}{4 \pi^2} \int d^2 k \int d^2 r \;
e^{-i \lambda \Delta( \vec r \, )-i \vec k \vec r} \;
\tilde a( \vec q + \vec k \, ) \tilde a( \vec p - \vec k \, )
 &=& 0 \; ,
\nn \\
\tilde a( \vec p \, ) \tilde a^{\dagger}( \vec q \, ) \; - \;
\frac{1}{4 \pi^2} \int d^2 k \int d^2 r
e^{i \lambda \Delta( \vec r \, ) - i \vec k \vec r} \;
\tilde a^{\dagger}( \vec q - \vec k \, ) \tilde a ( \vec p -
\vec k \, ) &=& \delta( \vec p - \vec q ) \; .              \label{17}
\eea
Let us use the notation
\be
\frac{1}{4 \pi^2} \int d^2 r
e^{i \lambda \Delta( \vec r \, ) - i \vec k \vec r} \;
= \; e^{\frac{i \, \lambda}{2 \, \pi} \CM ( \vec k \, ) } \; ,
                                                            \label{18}
\ee
where ~$\CM ( \vec k \, )$~ is the Fourier transform of
{}~$\Delta ( \vec r \, )$~:
\bea
\CM ( \vec k \, ) &=& \frac{1}{2 \pi} \int d^2 r
e^{- i \vec k \vec r} \: \Delta( \vec r \, )
\nn\\
&=& 2 \pi i \, \delta ( k_x ) \, \CP \frac{1}{ k_y} \; .    \label{19}
\eea
Then we find that
\be
e^{\frac{i \, \lambda}{2 \, \pi} \CM ( \vec k \, ) } =
\cos ( \pi \lambda ) \delta ( \vec k \, ) \, - \,
\frac{1}{\pi} \sin ( \pi \lambda ) \,
\delta ( k_x ) \, \CP \frac{1}{ k_y} \; .                   \label{20}
\ee
Hence, the q-commutation relations in momentum space transform into
\bea
\tilde a( \vec p \, ) \tilde a( \vec q \, ) \; - \;
\cos ( \pi \lambda ) \tilde a( \vec q \, ) \tilde a( \vec p \, )
\hspace{4cm} \nn\\
- \; \frac{1}{\pi} \sin (\pi \lambda ) \; \CP
\int_{- \infty}^{\infty} \: \frac{dk}{k} \;
\tilde a ( \vec q + \vec j k \, ) \tilde a ( \vec p - \vec j k \, )
 &=& 0 \; ,
\nn \\
\tilde a( \vec p \, ) \tilde a^{\dagger}( \vec q \, ) \; - \;
\cos ( \pi \lambda ) \tilde a^{\dagger}( \vec q \, )
\tilde a( \vec p \, ) \hspace{4cm} \nn\\
+ \; \frac{1}{\pi} \sin (\pi \lambda ) \; \CP
\int_{- \infty}^{\infty} \: \frac{dk}{k} \;
\tilde a^{\dagger}( \vec q - \vec j k \, )
\tilde a ( \vec p - \vec j k \, )
 &=& \delta( \vec p - \vec q ) \; .                         \label{21}
\eea
If ~$\sin( \pi \lambda ) \neq 0$~,there is a continuum of
contributions along the ~$k_{y}$~ momentum.This non-locality
in momentum space is a consequence of the ~$\vec r$~-dependence
in the grading factors in~(\ref{10})~ , i.e.~~$\Delta ( \vec r \, )
\neq const$~.

Closer inspection of Eq.~(\ref{21})~ at the point of coincidence,
i.e. ~$\vec p = \vec q$~,shows that the fermionic field
{}~$( \lambda = 1 \, mod \, 2 )$~ indeed satisfies the expected
hard-core condition. \\
For intermediate ~$\lambda \: , \: \sin ( \pi \lambda ) \neq 0$~,
a new unexpected "sum rule" emerges for the operator bilinears at the
same point in momentum space :
\be
\tilde a^2 ( \vec p \, ) ( 1 - \cos ( \pi \lambda )) \;
- \; \frac{\sin (\pi \lambda )}{\pi} \; \CP
\int_{- \infty}^{\infty} \: \frac{dk}{k} \;
\tilde a ( \vec p + \vec j k \, ) \tilde a ( \vec p - \vec j k \, )
= 0 \; .                                                    \label{22}
\ee
No hard-core condition appears directly , even if we start with
the q-anticommutator algebra in position space.

The connection between the momentum operators ~$a( \vec k \, )$~
and ~$\tilde a ( \vec k \, )$~ is
\bea
\tilde a ( \vec k \, ) &=& \frac{1}{2 \pi} \int \, d^2 r \,
e^{-i \vec k \vec r \, - \, i \omega( \vec r \, )} \:
\Psi ( \vec r \, )
\nn\\
&=& \int \, d^2 k' \, e^{- \frac{i}{2 \pi} \,
\tilde \omega ( \vec k \, - \, \vec k ' \, )} \: a( \vec k ' \, ) \; ,
                                                            \label{23}
\eea
where
\be
\tilde \omega ( \vec k \,) \, = \, - 2 \pi \lambda \: \tilde \theta
( \vec k \, ) \, \tilde \rho ( \vec k \, ) \hspace{1.5cm}   \label{24}
\ee
and
\bea
\tilde \theta ( \vec k \, ) &=& \frac{1}{2 \pi} \int d^2 r \:
e^{-i \vec k \vec r } \, \theta ( \vec r \, )
\nn\\
&=&  2 \pi \delta ( \vec k \, ) \, + \, \frac{1}{k^2} \,
 \tan \alpha \, +  \, \frac{1}{2} \, \CM ( \vec k \, ) \; , \label{25}
\\
\tilde \rho ( \vec k \, ) &=& \frac{1}{2 \pi} \int d^2 r \:
e^{-i \vec k \vec r } \, \rho ( \vec r \, )
\nn\\
&=& \frac{1}{2 \pi} \int \, d^2 k' \, \tilde a ^{\dagger}( \vec k ' )
\tilde a ( \vec k + \vec k ' \, ) \; .                      \label{26}
\eea
Here~$\alpha$~ is the polar angle between the ~$\vec k$~ and the
positive ~$k_{x}$~ axis , and ~$\CM ( \vec k \, )$~ is given by
Eq.~(\ref{19})~.
Hence
\bea
\tilde a ( \vec k \, ) &=& a( \vec k \, ) + i \lambda
\int \: d^2 k' \; \tilde{\theta} ( \vec k - \vec k ' \, ) \:
\tilde{\rho} ( \vec k - \vec k ' \, ) \: a( \vec k ' \, ) \;
\nn\\
& & + \frac{(i \lambda )^2}{2 \, !} \int d^2 k' d^2 k'' \;
[ \, \tilde{\theta} \, \tilde{\rho} \, ]
( \vec k - \vec k '' \, )
[ \, \tilde{\theta} \, \tilde{\rho} \, ]
( \vec k '' - \vec k ' \, ) \:
a( \vec k ' \, ) + \dots \; ,                               \label{27}
\eea
where we have formally expanded the exponential kernel in Eq.
{}~(\ref{23})~ in powers of the statistical parameter ~$\lambda$~.

Now we can use the conventional Fock-space method to construct
multiparticle states in momentum space by acting of the
{}~${\tilde a}^{\dagger}(\vec p \, )$~ operators on the vacuum.
For a two-particle state,for example,this yields
\be
\tilde a^{\dagger}( \vec p_{1} \, ) \tilde a^{\dagger}
( \vec p_{2} \, ) \; |0> \; = \;
\frac{1}{2 \pi} \int d^2 k \int d^2 r
e^{-i \lambda \theta ( \vec r \, ) - i \vec k \vec r} \;
a^{\dagger}( \vec p_{1} + \vec k \, )
a^{\dagger} ( \vec p_{2} -
\vec k \, ) \; |0> .                                        \label{39}
\ee
It is not obvious that this complicated structure,
in contrast to the corresponding structure given by
Eq.~(\ref{38})~ in position space,possesses the braiding
property.
However,braiding is still present.
To show this,let us mention that q-commutator algebras both
in position space Eqs.~(\ref{10})~,~(\ref{11})~ and in momentum
space Eqs.~(\ref{21})~ can be written in the ~$\cal R$-matrix
approach ~\cite{zach}:
\bea
\tilde{\Psi}( \vec r_{1} \, ) \tilde {\Psi}( \vec r_{2} \, ) \; - \;
\int d^2 x \int d^2 y \;
{\cal R} ( \vec r_{1}, \vec r_{2}, \vec x , \vec y \, ) \:
\tilde{\Psi}( \vec y \, ) \tilde{\Psi}( \vec x \, )
 &=& 0 \; ,
\nn \\
\tilde {\Psi}( \vec r_{1} \, ) \tilde {\Psi}^{\dagger}
( \vec r_{2} \, ) \; - \;
\int d^2 x \int d^2 y \;
{\cal R} ( \vec y , \vec r_{1}, \vec r_{2}, \vec x \, ) \:
\tilde{\Psi}^{\dagger}( \vec y \, ) \tilde{\Psi}( \vec x \, )
&=& \delta ( \vec r_{1} - \vec r_{2} ) \; .                  \label{40}
\eea
The ~$\cal R$~-matrix in position space is given by
\be
{\cal R} ( \vec r_{1}, \vec r_{2}, \vec r_{3}, \vec r_{4} \, ) \: = \:
e^{-i \lambda \Delta( \vec r_{1} - \vec r_{2} \, )} \:
\delta ( \vec r_{1} - \vec r_{3} ) \,
\delta ( \vec r_{2} - \vec r_{4} )                          \label{41}
\ee
and the ~$\cal R$~-matrix in momentum space is given by
\bea
&&{\cal R} ( \vec k_{1}, \vec k_{2}, \vec k_{3}, \vec k_{4} \, ) \:
= \: \delta ( \vec k_{3} + \vec k_{4} - \vec k_{1} - \vec k_{2} \, ) \:
\nn \\
&& \; \; \; \times
[ \cos ( \pi \lambda ) \delta ( \vec k_{1} - \vec k_{3} \, ) \, + \,
\frac{1}{\pi} \sin ( \pi \lambda ) \,
\delta ( k_{3x} - k_{1x} ) \,
\CP \frac{1}{k_{3y} - k_{1y} } \, ] \; .                    \label{42}
\eea
In order that the above ~$\cal R$~-matrix algebra be associative,
the following conditions have to be satisfied:

(i) Yang-Baxter equation:
\bea
&&\int \int \int d^2 x  d^2 y d^2 z \;
{\cal R} ( \vec r_{1}, \vec r_{2}, \vec x , \vec y \, ) \,
{\cal R} ( \vec y , \vec z , \vec r_{3}, \vec r_{4} \, ) \,
{\cal R} ( \vec x , \vec r_{5}, \vec r_{6}, \vec z \, ) \,
\nn \\
&& \; \; \; = \, \int \int \int d^2 x  d^2 y d^2 z \;
{\cal R} ( \vec r_{2}, \vec r_{5}, \vec x , \vec y \, ) \,
{\cal R} ( \vec z , \vec x , \vec r_{6}, \vec r_{3} \, ) \,
{\cal R} ( \vec r_{1} , \vec y , \vec z, \vec r_{4} \, ) \; ,
\eea
(ii) hermiticity:
\be
{\cal R} ( \vec r_{1}, \vec r_{2},\vec r_{3}, \vec r_{4} \, ) \: = \:
{\cal R}^{\ast} ( \vec r_{4}, \vec r_{3},\vec r_{2}, \vec r_{1} \, )
\; ,
\ee
(iii) Hecke condition
\be
\int \int  d^2 x  d^2 y  \;
\check {\cal R} ( \vec r_{1}, \vec r_{2}, \vec x , \vec y \, ) \,
\check {\cal R} ( \vec x , \vec y , \vec r_{3}, \vec r_{4} \, ) \, = \,
\delta ( \vec r_{1} - \vec r_{3} \, ) \:
\delta ( \vec r_{2} - \vec r_{4} \, ) \; ,
\ee
or symbolically:
\bea
( \check{\cal R} \, - \, 1 ) \, ( \check{\cal R} \, + \, 1 )
\: = \: 0 \, ,
\nn
\eea
where
\bea
\check{\cal R} \: &=& \: P \, {\cal R} \; \; \; \; \; \; \; and
\nn \\
P ( \vec r_{1}, \vec r_{2},\vec r_{3}, \vec r_{4} \, ) \: &=& \:
\delta ( \vec r_{1} - \vec r_{4} \, ) \:
\delta ( \vec r_{2} - \vec r_{3} \, ) \; .
\nn
\eea
It is straightforward to show that the ~$\cal R$~-matrices
{}~(\ref{41})~and~(\ref{42})~ satisfy all the above
conditions both in position and momentum spaces.
Hence this proves the braiding properties both in
position and momentum spaces.

In order to analyze the origin of the q-deformed algebra~(\ref{10})~,
we introduce a new field ~$\Phi ( \vec r \, )$~ defined as
\be
\Phi ( \vec r \, ) = e^{-i \, \omega ' ( \vec r \, )}
\Psi ( \vec r \, ) \; ,                                     \label{30}
\ee
with
\be
\omega ' ( \vec r \, ) = - \lambda \, \int d^2 r' \:
\arctan \frac{y-y'}{x-x'} \; \rho ( \vec r ' \, ) \; .      \label{31}
\ee
Then,
\be
\vec {\nabla} \omega ' ( \vec r \, ) = \vec A ( \vec r \, ) \,
- \, \vec A_{res} ' ( \vec r \, ) \; ,                      \label{32}
\ee
where
\be
\vec A_{res} ' ( \vec r \, ) \, = \, \pi \lambda \, \vec i \,
\int_{- \infty}^{\infty} \, dy' \sgn (y-y') \, \rho (x,y') \: .
                                                            \label{33}
\ee
The ~$\vec A_{res} '$~ potential originates from the cut along
the y-axis.

The field~$\Phi ( \vec r \, )$~ remains bosonic,satisfying
the same commutation relations as ~$\Psi ( \vec r \, )$~ in Eq.
{}~(\ref{2})~.This follows from the fact that ~$\arctan \frac{y}{x}$~
is a symmetric function under the inversion
{}~$\vec r \, \rightarrow \, - \vec r$~.Hence,
{}~$\Delta( \vec r \, ) = 0$~.The Hamiltonian of the new bosonic
field ~$\Phi ( \vec r \, )$~ is not free because of~$\vec A_{res} '$~.
Note that~$\vec B = \vec \nabla \times \vec A_{res} ' =
2 \pi \lambda \: \hat n \, \rho ( \vec r \, )$~,~i.e.~ the
magnetic field does not change under the transformation~(\ref{30})~.
Hence all the physical consequences are the same as for the initial
non-relativistic Chern-Simons theory.

Finally,we connect the q-deformed anyonic field~$\tilde \Psi
( \vec r \, )$~ with the bosonic field~$\Phi ( \vec r \, )$~:
\be
\tilde \Psi ( \vec r \, ) = e^{-i \, [ \, \omega ( \vec r \, ) \,
- \, \omega ' ( \vec r \, ) \, ]} \: \Phi ( \vec r \, ) \; ,
                                                            \label{34}
\ee
where~$\omega ( \vec r \, )$~ and ~$ \omega ' ( \vec r \, )$~ are
given by ~(\ref{6})~and~(\ref{31})~,respectively.

The effective kernel in the expression~$\omega ( \vec r \, ) \,
- \, \omega ' ( \vec r \, )$~ is
\bea
\theta_{eff}( \vec r \, ) \; &=& \;
\theta ( \vec r \, ) \; - \; \arctan ( \, \frac{y}{x} \, )
\nn \\
&=& \left \{ \begin{array}{ll}
             \pi - \frac{\pi}{2} \sgn (xy) - \frac{\pi}{2} \sgn (y) &
              {\rm if} \; \; y \neq 0  \\
             \frac{\pi}{2} ( \,1 \, - \, \sgn (x) \, ) &
              {\rm if} \; \; y = 0 \; .                     \label{35}
             \end{array}
    \right.
\eea
It is important to note that only the antisymmetric part of
{}~$\theta_{eff}$~ deforms the commutation relations,yielding
{}~$\Delta ( \vec r \, )$~ given by Eq.~(\ref{11})~.The effective
kernel ~$\theta_{eff}$~changes the cut along the positive x-axis
into the cut along the y-axis, and vice versa.Hence~$\theta_{eff}$~
relates ~$\vec A_{res}$~to~$\vec A_{res} '$~.

The origin of the transmutation from bosons to anyons ( or
fermions to \\ anyons ) is in the antisymmetric part of the angle
function.This suggests a generalization of the above construction
for a two-dimensional space to a space ~${\CR}^{n}$~ with an arbitrary
number of dimensions n .\\
Let ~$x \, = \, ( x_{1} , \dots x_{n} ) \, \in \, {\CR}^{n}$~,
then we define
\bea
\theta_{eff}( \,x\, - \, x' \, ) \; &=& \;
 \left \{ \begin{array}{ll}
           \sgn ( x_{n} \, - \, x_{n} ') &
           {\rm if} \; \; x_{n} \, \neq \, x_{n} ' \\
           \sgn ( x_{n-1} \, - \, x_{n-1} ') &
           {\rm if} \; \; x_{n} \, = \, x_{n} ' \\
           \; \; \vdots & \\
           \sgn ( x_{1} \, - \, x_{1} ') &
           {\rm if} \; \; x_{n} \, = \, x_{n} ' , \; \; \dots \\
               & \: \; \; \; \; x_{2} \, = \, x_{2} ' \\
           \hspace{1.4cm}  0 & {\rm if} \: \; \; \; x \, = \, x ' \; .
           \end{array}                                      \label{36}
    \right.
\eea
The interpolation between bosons and fermions is achieved
by ~$\tilde \Psi ( \vec r \, )$~ given by Eqs.~(\ref{5})~,
{}~(\ref{6})~ and ~(\ref{36})~.

The N-particle anyonic wave function for
{}~$\vec r_{1} \neq \vec r_{2} \neq \dots \neq \vec r_{N}$~
has a simple form expressed in terms of the bosonic
field ~$\Psi (\vec r \, )$~,
\bea
&& \tilde{\Psi}^{\dagger} (\vec r_{i_{1}} ) \:
\tilde{\Psi}^{\dagger}(\vec r_{i_{2}})
\dots \tilde{\Psi}^{\dagger} (\vec r_{i_{N}} ) \; |0> =
\nn \\
&& \; \; \; = \,
e^{-i \, \lambda \, \pi{\Sigma}_{\alpha < \beta }
\theta_{eff} ( \vec r_{i_{\alpha} } - \vec r_{i_{\beta} } )}
\: {\Psi}^{\dagger} (\vec r_{1} ) \: {\Psi}^{\dagger}(\vec r_{2})
\dots {\Psi}^{\dagger} (\vec r_{N} )
 \; |0>
\nn \\
&& \; \; \; = \,
e^{i \, \lambda \, 2 \pi [ \frac{N(N-1)}{4} - P(i_{1} ,
\dots i_{N})]}
\: {\Psi}^{\dagger} (\vec r_{1} ) \: {\Psi}^{\dagger}(\vec r_{2})
\dots {\Psi}^{\dagger} (\vec r_{N} )
 \; |0>
\eea
Here ~$i_{1}, \dots i_{N}$~ is the permutation of indices
{}~$1,2, \dots N$~, and ~$P(i_{1}, \dots i_{N})$~ denotes the
number of inversions in permutation ~$i_{1}, \dots i_{N}$~
in respect to the normal order ~$1,2, \dots N$~,defined by
{}~$\theta_{eff} (\vec r_{\alpha} - \vec r_{\beta})= +1$~
iff ~$\alpha > \beta$~.The braiding properties hold both in
position and momentum space,since the corresponding ~$\cal R$~-matrices
satisfy the Yang-Baxter equation and the Hecke condition.
The properties of the free Hamiltonian with such a
q-deformed field are under investigation.
\bea
\vspace{2cm}
\nn
\eea
{\large \bf Acknowledgment} \\
\vspace{1cm}
This work was supported by the Scientific Fund of the Republic of
Croatia.

\end{document}